\documentstyle[12pt]{article}
\begin{document}
\baselineskip=24pt
\pagestyle{plain}

\begin{center}
{\Large \bf Canonical Theory of 2+1 Gravity}
\\
\vspace{1cm}
{\large M.Kenmoku 
\footnote[1]{Talk given at LLWI-2000: 
From Particles to Universe, Alberta, 20-26 February 2000.
\\
\hspace{2mm}
\footnotemark[1]
kenmoku@phys.nara-wu.ac.jp}, 
T.Matsuyama \footnote[2]{matsuyat@nara-edu.ac.jp}, 
R.Sato \footnote[3]{reika@phys.nara-wu.ac.jp}
and S.Uchida \footnote[4]{satoko@phys.nara-wu.ac.jp}}
\\
\vspace{0.5cm}
{\small \it \footnotemark[1] 
Department of Physics, Nara Women's University, 
Nara 630-8506, Japan \\ 
\footnotemark[2]
Department of Physics, Nara University of Education, Takabatake-cho, 
Nara 630-8528, Japan \\
\footnotemark[3] \footnotemark[4]
Graduate School of Human Culture, Nara Women's University, 
Nara 630-8506, Japan}
\end{center}

\vspace{0.5cm}

Recently 2+1 dimensional gravity theory, especially ${\rm AdS_3}$ 
has been studied extensively \cite{bib:djt84,bib:btz92}. 
It was shown to be equivalent to 
the 2+1 Chern-Simon theory \cite{bib:witten88}
and has been investigated to understand 
the black hole thermodynamics, i.e. Hawking temperature \cite{bib:hawking75}
and others. 
The purpose of this report is to investigate the canonical formalism of 
the original 2+1 Einstein gravity theory instead of the Chern-Simon theory. 
For the spherically symmetric space-time, 
local conserved quantities(local mass and angular momentum) are 
introduced and using them canonical quantum theory is defined. 
Constraints are imposed on state vectors and solved analytically.
The strategy to obtain the solution is followed by 
our previous work \cite{bib:kkty99} .

\section{Canonical formalism}

We start to consider the Einstein-Hilbert action with cosmological constant
\( \lambda \)  in 2+1 dimensional space-time,
\begin{equation}
I={1 \over 16 \pi G_2} \int d^{3}x \sqrt{-^{(3)}g} (^{(3)}R-2\lambda) .
\label{action}
\end{equation}
The gravitational constant in 2+1 dimension is set to 
$G_2=1/4$ in the following. 
The metrics in polar coordinate are expressed in ADM 
decomposition \cite{bib:adm62} as
\begin{eqnarray}
ds^{2}=-N^{2}dt^{2}+\Lambda^{2}(dr+N^{r}dt)^{2}+R^{2}(d\phi+N^{\phi}dt)^{2}
\\ \nonumber
+2C(dr+N^{r}dt)(d\phi+N^{\phi}dt) ,
\label{metric}
\end{eqnarray}
where all metrics are assumed to be function of time $t$ and radial 
coordinate $r$. In the following, dot and dash denotes the derivative 
with respect to $t$ and $r$.  

The action in canonical formalism is in the form 
\begin{eqnarray}
I &=&  \int dt \, dr \;[P_{\Lambda}\dot{\Lambda}+P_{R}\dot{R}+P_{C}\dot{C}
-(NH + N^{r} H_{r} + N^{\phi} H_{\phi})] 
\nonumber \\
& & -  \int dt \, dr \biggl( [(\Lambda P_{\Lambda} + C P_{C})N^{r} ]^{\prime} 
+  [ ({C \over \Lambda} P_{\Lambda} + R^{2} P_{C}) N^{\phi} ]^{\prime} \biggl) \;,
\end{eqnarray}
where canonical momenta are
\begin{eqnarray}
P_{\Lambda} &=& {\partial {\cal L} \over \partial \dot{\Lambda}}=
{2 \Lambda R (N^{r} R^{\prime}-\dot{R}) \over N \sqrt{h}} \;,
\\
P_{R} &=& {\partial {\cal L} \over \partial \dot{R}}=
{2R[C{N^{\phi}}^{\prime}+\Lambda \{(\Lambda N^{r})^{\prime} - \dot{\Lambda} \}] 
\over N \sqrt{h}} \;,
\\
P_{C} &=& {\partial {\cal L} \over \partial \dot{C}}=
-{(N^{r}C)^{\prime} + R^{2} {N^{\phi}}^{\prime} - \dot{C} \over 
N \sqrt{h} } \;,
\end{eqnarray}
and the Hamiltonian and the momentum constraints are defined as 
\begin{eqnarray}
H &=& - {\sqrt{h} \over 2} \biggl( {P_{\Lambda}P_{R} \over \Lambda R} - {P_{C}}^{2} \biggl)
-2 \biggl( - {{R^{\prime}}^{2}+R R^{\prime \prime} \over \sqrt{h}}+
{R R^{\prime} h^{\prime} \over 2h\sqrt{h}} \biggl)-2 \lambda \sqrt{h} \;,
\\
H_{r}&=& P_{R} R^{\prime} -C {P_{C}}^{\prime} -\Lambda {P_{\Lambda}}^{\prime} \;,
\\
H_{\phi}&=& - \biggl({C \over \Lambda} P_{\Lambda}+R^{2} P_{C} \biggl)^{\prime} \;.
\end{eqnarray}

It is essential to introduce the local conservation quantities, 
the angular momentum \( J \) and the mass function \( M \) 
as follows.
\begin{eqnarray}
J &:=& - \int dr H_{\phi} = {C \over \Lambda} P_{\Lambda}+R^{2} P_{C} \;,
\label{angular} \\
M &:=& 
- \int dr \biggl( {R R^{\prime} \over \sqrt{h} }H + {P_{\Lambda} \over \Lambda}H_{r}
+ P_{C} H_{\phi} \biggl) \nonumber
\\ &=& 
{1 \over 2} \biggl({P_{\Lambda}}^{2} + {2C P_{\Lambda} P_{C} \over \Lambda}
+ R^{2} {P_{C}}^{2} - { (R R^{\prime})^{2} \over h } - \lambda R^{2} \biggl) 
\label{mass}  
\end{eqnarray}

We make transformation from old variables \( \Lambda, R\) and \( C \) into 
new variables  
\begin{equation}
\left( \begin{array}{c} \Lambda \\ R \\ C \end{array} \right)
\longrightarrow
\left( \begin{array}{c} \ \bar{\Lambda} \\ \bar{R} \\ \bar{C} \end{array} \right)
= \left( \begin{array}{c} \sqrt{\Lambda^{2}-{C^{2} R^{-2}}} \\ R 
\\ {C  R^{-2}} \end{array} \right) . 
\label{pointra}
\end{equation} 
The corresponding momenta are transformed as
\begin{eqnarray}
\left( \begin{array}{c} P_{\Lambda} \\ P_{R} \\ P_{C} \end{array} \right)
\longrightarrow
\left( \begin{array}{c} P_{\bar{\Lambda}} \\ P_{\bar{R}} \\ P_{\bar{C}}
\end{array} \right)
= \left( \begin{array}{c} \bar{\Lambda} \Lambda^{-1} P_{\Lambda} \\
{C^{2} \Lambda^{-1} R^{-3}} P_{\Lambda} + P_{R} + 2C R^{-1} P_{C}
\\ C \Lambda^{-1} P_{\Lambda} + R^{2} P_{C} \end{array} \right) .
\label{canmom}
\end{eqnarray} 

\section{Quantum solutions}
Next we proceed the quantum theory in the Schr${\rm \ddot{o}}$dinger picture 
and the quantized operators are denoted by the notation hat. 
Our strategy is to solve the eigenvalue equation 
for $\hat{J}$, $\hat{M}$ and the constraint equation 
for $H_r$ step by step instead of solving the constraint equations 
\( \hat{H} \Psi=0, \hat{H_{r}} \Psi=0 \) and \( \hat{H_{\phi}} \Psi=0 \). 
\\
Step 1: Angular momentum eigen equation \\
The eigenvalue equation of the local angular momentam 
(Eq. (\ref{angular}))
\begin{eqnarray}
\hat{J} \Psi = \hat{P_{\bar{C}}} \Psi = j \Psi \;
\end{eqnarray}
is solved with the eigenvalue $j$ and 
the eigen function is obtained in the form 
\begin{eqnarray}
\Psi = e^{ij \Phi} u(\bar{\Lambda},\bar{R}) \;,
\end{eqnarray}
with
\begin{eqnarray}
\Phi = \int dr \bar{C}(r) \;.
\end{eqnarray}
\\
\noindent
Step 2: Momentum constraint equation \\ 
The radial momentum constraint equation 
\begin{eqnarray}
\hat{H}_{r} \Psi = 
 (\bar{R}^{\prime} \hat{P}_{\bar{R}} - 
 \bar{\Lambda} (\hat{P}_{\bar{\Lambda}})^{\prime} ) 
 e^{ij\Phi} u(\bar{\Lambda},\bar{R}) 
= 0 \;,
\end{eqnarray}
restricts the functional form of the wave function as 
\begin{equation}
\Psi = e^{ij \Phi} u(Z) \;,
\end{equation}
where we introduce variable \( Z \)
\begin{eqnarray}
Z = \int dr \bar{\Lambda} f(\bar{R},\chi)  
  = \int dr \int^{\bar{\Lambda}(r)} d\bar{\Lambda} 
                  \bar{f} (\bar{R},\chi) ,
                  \label{zett}
\end{eqnarray}
with 
\begin{eqnarray}
\chi := {{R^{\prime}}^{2} \bar{\Lambda}^{-2}} \;.
\label{chi}
\end{eqnarray}
The arbitraly function $f$ and $\bar{f}$ are related each other: 
\begin{eqnarray}
f(\bar{R},\chi) &=& -\int^{\chi} d\chi {\bar{f}(\bar{R},\chi) \over 2\chi} \;.
\end{eqnarray}
\\
\nonumber 
Step 3:  Mass eigen equation \\
The local mass operator \( \hat{M} \) is defined as 
\begin{equation}
\hat{M} -m = {1 \over 2} 
            A \hat{P}_{\bar{\Lambda}}A^{-1} \hat{P}_{\bar{\Lambda}}
        + {1 \over 2} ( -\chi + \hat{F}(\bar{R})) \;,
\end{equation}
where
\begin{eqnarray}
\hat{F}(\bar{R}) =  1-2m-\lambda \bar{R}^{2} + 
                    {1 \over 4} {\hat{J}^{2} \bar{R}^{-2}} \;, 
\end{eqnarray}
and
\begin{equation}
A = A_{Z}(Z) \bar{A}(\bar{R},\chi) \;,
\end{equation}
which is called ordering factor.
We take $\bar{A}$ as 
\begin{equation}
\bar{A} = {\delta Z \over \delta \bar{\Lambda}} = \bar{f} 
        = \sqrt{\chi - F_j(\bar{R})} \;,
\label{fbar}
\end{equation}
where 
\begin{equation}
F_j(\bar R):= \hat{F}\mid_{\hat{J}=j} \label{fj} (\bar R)\ .
\end{equation}
Then using the mass operator for each eigenvalue of angular momentum $j$ 
\begin{equation}
\hat{M}_{j}:= \hat{M}\mid_{\hat{J}=j} \ ,
\end{equation}   
the mass eigen equation 
\begin{equation}
\hat{M}_{j} u_{j,m}(Z)= m u_{j,m}(Z) \; ,
\end{equation}
can reduce to the equation with respect to $Z$
\begin{equation}
{d^{2}u_{j,m}(Z) \over dZ^{2}} 
- {A_{Z}}^{-1} {\delta A_{Z} \over \delta Z}{d u_{j,m}(Z) \over dZ} 
+ u_{j,m}(Z) = 0 .
\end{equation} 
If we choose the remaining ordering factor as \( A_{Z}=Z^{2\nu-1} \), 
the above equation becomes the Bessel equation 
\begin{equation}
{d^{2}u_{j,m}(Z) \over dZ^{2}} - {2\nu-1 \over Z}{du_{j,m}(Z) \over dZ} 
+ u_{j,m}(Z) = 0 ,
\end{equation}
and the solution is
\begin{equation}
u^{(\nu)}_{j,m}(Z)=Z^{\nu}[b_{1}H_{\nu}^{(1)}(Z)+b_{2}H_{\nu}^{(2)}(Z)] ,
\label{unuz}
\end{equation}
where \( H_{\nu}(Z) \) is the Hankel function.

\section{Summary}

In conclusion, the general form of quantum wave function is obtained 
\begin{equation}
\Psi (Z)= \sum_{j,m} c_{j,m} e^{ij \Phi} u^{(\nu)}_{j,m}(Z) \;,
\label{psi}
\end{equation}
where $c_{j,m}$ are the expansion coefficients, 
\( u^{(\nu)}_{j,m}(Z) \) is expressed by 
Eq. (\ref{unuz}) 
and \(Z \) is expressed using Eqs. (\ref{zett}) and (\ref{fbar}) as
\begin{eqnarray}
Z &=& \int  dr \int ^{\bar{\Lambda} (r)} 
       d \bar{\Lambda} \sqrt{ \chi - F_j(\bar{R}) } 
       \nonumber \\
  &=& \int dr \biggl(\bar{\Lambda} \sqrt{\chi - F_j(\bar{R})} 
     -  {\bar{R}}^{\prime} 
\ln \biggl| { \sqrt{\chi} + \sqrt{\chi - F_j(\bar{R})}
        \over \sqrt{\mid F_j(\bar{R}) \mid}}\biggl| 
\biggl) \;,
\end{eqnarray}
where $\chi$ and $F_j$ are given in Eqs. (\ref{chi}) and (\ref{fj}). 
It is worthwhile to note that the analytic solution in Eq.(\ref{psi}) is shown 
to satisfy the original constraint equations as well as 
the Wheeler-DeWitt equation. 
Therefore we have successfully obtained the analytic solution for 
the Wheeler-DeWitt equation. The interpretation for the wave function 
will be appeared in separate paper.

\end{document}